\documentclass[journal]{IEEEtran}

\usepackage{graphics} 
\usepackage{epsfig} 
\usepackage{mathptmx} 
\usepackage{amsthm}
\usepackage{amsmath} 
\usepackage{amssymb}  
\usepackage{cuted}
\usepackage{float}
\usepackage{color}
\usepackage{url}
\usepackage{bm}

\usepackage[ruled,vlined,linesnumbered]{algorithm2e} 
\usepackage{tabularx}
\usepackage{caption}
\usepackage{subcaption}

\newtheorem{definition}{Definition}

\newtheorem{example}{Example}

\makeatletter 
\def\ps@IEEEtitlepagestyle{%
  \def\@oddfoot{\mycopyrightnotice}%
  \def\@evenfoot{}%
}
\def\mycopyrightnotice{%
  \begin{minipage}{\textwidth}
  \centering \scriptsize
  Copyright~\copyright~2026 IEEE. Personal use of this material is permitted. Permission from IEEE must be obtained for all other uses, in any current or future media, including reprinting/republishing this material for advertising or promotional purposes, creating new collective works, for resale or redistribution to servers or lists, or reuse of any copyrighted component of this work in other works. 
  \end{minipage}
}
\makeatother 

\begin{document}
%
\title{
Toward Secure Operation and Management  (O\&M) of Satellite Constellations:
Efficiency, Resilience, and Reliability in a Network Perspective
} 


%
\author{
Linan Huang, 
Peilong Liu, 
Xi Chen, 
Zhiyuan Lin, 
Jian Yan\IEEEauthorrefmark{2}, 
and Linling Kuang\IEEEauthorrefmark{2}%
\thanks{L. Huang, P. Liu, X. Chen, Z. Lin, and J. Yan are with the Beijing National Research Center for Information Science and Technology (BNRist), Tsinghua University, Beijing 100084, China (e-mail: huanglinan@tsinghua.edu.cn; plliu@tsinghua.edu.cn; chenxiee@tsinghua.edu.cn; zy-lin@tsinghua.edu.cn; yanjian\_ee@tsinghua.edu.cn).}%
\thanks{L. Kuang is with the Beijing National Research Center for Information Science and Technology, and the State Key Laboratory of Space Network and Communications, Tsinghua University, Beijing 100084, China (e-mail: kll@tsinghua.edu.cn).}
\thanks{\IEEEauthorrefmark{2}Corresponding authors: Jian Yan and Linling Kuang.}%
}
\markboth{Journal of \LaTeX\ Class Files,~Vol.~14, No.~8, August~2015}%
{Shell \MakeLowercase{\textit{et al.}}: Bare Demo of IEEEtran.cls for IEEE Journals}
%



\maketitle

\begin{abstract}
Satellite constellations equipped with Inter-Satellite Links (ISLs) and onboard packet switching enable real-time Operation and Management (O\&M) across globally distributed satellites, but at the same time broaden the attack surface and expose the system to unprecedented cybersecurity threats. 
Existing efforts have primarily focused on optimizing cryptography for single-satellite, point-to-point links, without considering \textcolor{black}{how a broader notion of security should be applied at the constellation level}. 
To bridge this gap, this article advances in two directions:  a \textit{\textcolor{black}{network} perspective}—from individual satellites to constellation-wide architectures; and an \textit{expanded security perspective}—from isolated cryptography to system-level security that embeds efficiency, resilience, and reliability. 
Together, these two expansions motivate three fundamental questions: (i) how can \textit{efficient} security mechanisms be designed for dynamic constellation topologies that require adaptive \textcolor{black}{onboard} routing; (ii) how can a constellation O\&M \textit{resiliently} recover to an acceptable operational state \textcolor{black}{under worst-case failures of onboard security functions}; and (iii) how can the \textit{reliability} of onboard security functions be enhanced under stringent \textcolor{black}{onboard} resource constraints. 
To address these challenges, we first construct a constellation-wide \textcolor{black}{hybrid} security framework that safeguards \textcolor{black}{content field with high semantic sensitivity} using End-to-End (E2E) encryption, while protecting routing-related fields through Moving Target Defense (MTD). 
Next, we introduce a ciphered-mode and safe-mode management mechanism with an $M$-delayed fallback that balances recovery timeliness against exploitability. 
Finally, to enhance reliability \textcolor{black}{of onboard security functions}, we propose \textcolor{black}{security-aware routers that manage plaintext/ciphered modes and orchestrate access to a shared pool of onboard cipher modules}, thereby enabling redundancy to be shared across multiple endpoints and significantly extending the duration of secure operation in ciphered mode. 
\textcolor{black}{Overall, these solutions comply with existing standards defined by standardization organizations, including Digital Video Broadcasting (DVB) and the Consultative Committee for Space Data Systems (CCSDS), while translating their conceptual security principles into more practical, system-level mechanisms}. 
\end{abstract}
 

\section{Introduction} 
\label{sec:Intro}
Satellite communications are evolving from single-satellite systems to interlinked constellations with onboard packet switching. 
Such architectures enable seamless, all-time global control from a single ground station and deliver real-time connectivity across space, air, sea, and land \cite{kodheli2020satellite}. 
Yet, the very features that extend their reach and utility also amplify system complexity, widening the attack surface across user, space, ground, and link segments \cite{salim2024cybersecurity}. This exposes \textcolor{black}{satellite} constellations to risks such as signaling hijacking, terminal impersonation, and cross-domain attacks \cite{huang2025consolidated}.


\begin{figure}[t]
\centering
\includegraphics[width=1 \columnwidth]{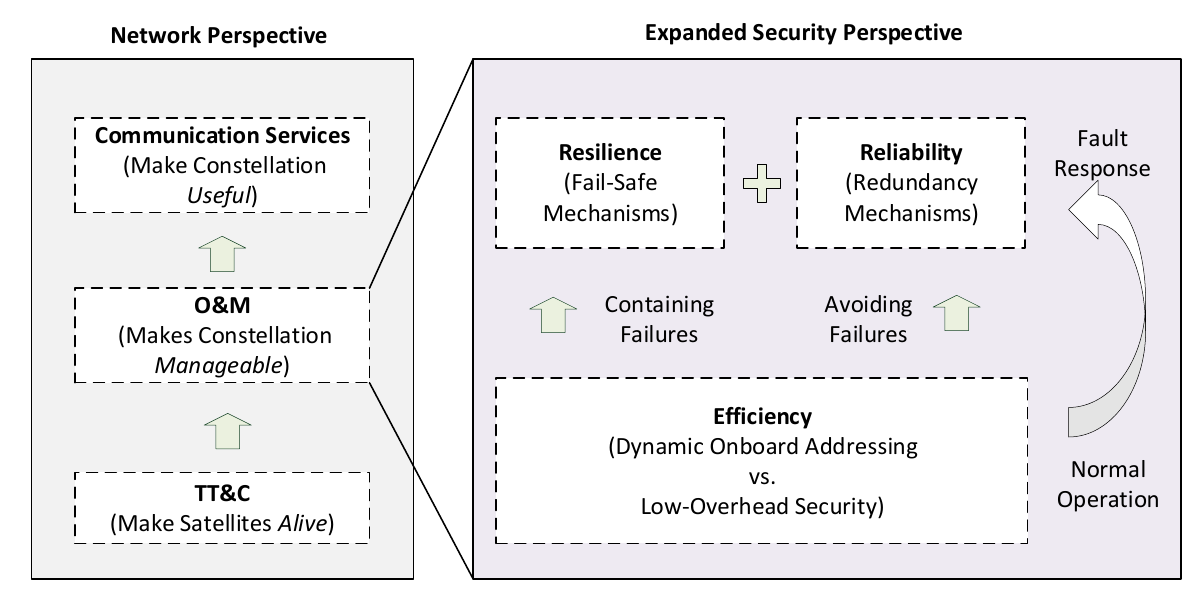}
\caption{ 
\textcolor{black}{
Overall structure of the paper, which applies an expanded security perspective to the network perspective of a satellite constellation.
Focusing on constellation O\&M, the paper first addresses security issues under \textit{normal operation}, with particular emphasis on dynamic constellation characteristics and the need to maintain operational efficiency.
It then considers \textit{fault response} mechanisms, which are categorized into resilience and reliability, to handle failures arising from the newly introduced security mechanisms. 
}
}
\label{fig:PaperStructure}
\end{figure}

\textcolor{black}{As illustrated in the left part of Fig. \ref{fig:PaperStructure}, a networked satellite constellation can be conceptually organized into three functional layers.
At the bottom layer, Telemetry, Tracking, and Command (TT\&C) ensures the survivability and basic controllability of each satellite platform.
The middle layer of Operation and Management (O\&M) performs constellation-wide management of both satellite platforms and payloads, including configuration, coordination, and resource allocation to support service operation. 
Supported by the TT\&C and O\&M layers, the constellation ultimately provides communication services to external users.} 
This paper focuses on safeguarding constellation O\&M, given their importance to both satellite survival in orbit and service continuity. 
In particular, \textcolor{black}{we examine both natural threats and adversarial threats}, with the objective of preserving the confidentiality, integrity, and availability (CIA) triad.

Cryptography lays the foundation for achieving the CIA triad, mainly through functions such as authentication, key agreement, and encryption. 
To adapt cryptography for deployment in satellite systems, researchers have focused on the unique challenges of space systems: addressing the extra encryption overhead that squeezes the scarce Space-to-Ground Link (SGL) bandwidth; selecting ciphers with high scores in throughput, memory footprint efficiency, security, and energy efficiency \cite{saha2019ensuring} to cope with strict Size, Weight, and Power (SWaP) limits onboard; developing algorithms resilient to long delays and bursty errors in satellite channels;  and ensuring smooth compatibility with terrestrial network protocols \cite{roy2005security}. 
Despite these advances, critical gaps persist. Existing efforts have primarily focused on single-satellite, point-to-point encryption, while constellations demand efficient system-wide protection. Moreover, while cryptography has been realized under nominal conditions, far less has been done to ensure O\&M resilience and cipher module reliability when the harsh space environment \textcolor{black}{or advanced attacks induce} anomalies.

To this end, we identify three key challenges \textcolor{black}{arising from the application of an expanded security perspective, as illustrated in the right part of Fig. \ref{fig:PaperStructure}. 
The first challenge concerns normal operation, where protecting across-satellite O\&M data frames must be reconciled with dynamic onboard routing under changing constellation topologies and strict efficiency requirements.
In particular, two classical security mechanisms cannot be directly applied: End-to-End (E2E) encryption of all fields impedes routing functionality, while Hop-by-Hop (H2H) encryption introduces prohibitive overhead. 
The second and third challenges of resilience and reliability arise when the newly introduced security mechanisms do not function as intended, a critical concern in satellite systems where on-orbit components are not repairable. 
Resilience focuses on handling worst-case outcomes caused by all potential abnormalities of the security mechanisms and on designing fail-safe strategies to contain their impact.} 
For example, a satellite operating in an abnormal ciphered mode may fail to decrypt telecommands, thereby rendering the system inoperable. 
The key challenge is to design fallback mechanisms that are both timely (i.e., ensuring recoverability when security functions fail) and non-exploitable by attackers during normal operation.
\textcolor{black}{Reliability, on the other hand, aims to prevent failures of security functions by introducing redundancy-based mechanisms.} 
Under strict SWaP constraints, the key challenge is to extend the Mean Time to Failure (MTTF) of onboard security functions without additional backup modules. 

This article aims to fill these gaps. 
For the first challenge of efficient constellation-wide protection architectures, we introduce the concept of Moving Target Defense (MTD) \cite{sengupta2020survey}, \textcolor{black}{a proactive defense approach that improves security by continuously changing system configurations over time, thereby limiting an attacker’s ability to exploit static system knowledge. In particular, we classify O\&M data frame fields according to semantic sensitivity and processing efficiency, and propose a hybrid protection scheme that applies E2E encryption to content fields while protecting routing-related fields using MTD.} 
For the second challenge of O\&M resilience, we design a ciphered-mode and safe-mode management mechanism. By introducing a delayed switching window of size $M$ and a detection module to suppress unintended transitions, this mechanism provides a principled balance between timely recovery from module faults and resistance to adversarial exploitation. 
For the third challenge of cipher reliability, we introduce \textcolor{black}{a security-aware onboard router structure with} the Unified and Pooled Onboard Cipher Module (UP-OCM) to implement invocation-based encryption. 
By consolidating cipher units into a shared pool coordinated by the onboard router, UP-OCM enables scalable redundancy sharing across O\&M endpoints and significantly prolongs the duration of normal ciphered-mode operation. 

The rest of the article is organized as follows. Section \ref{sec:System Architecture} introduces the system architecture and security goals. Sections \ref{sec:1}, \ref{sec:2}, and \ref{sec:3} elaborate on our proposed solutions to the three challenges. Finally, Section \ref{sec:conclusion} concludes the article.

\section{\textcolor{black}{System Architecture and Threat Model}}
\label{sec:System Architecture}

\begin{figure*}[t]
\centering
\includegraphics[width=1 \linewidth]{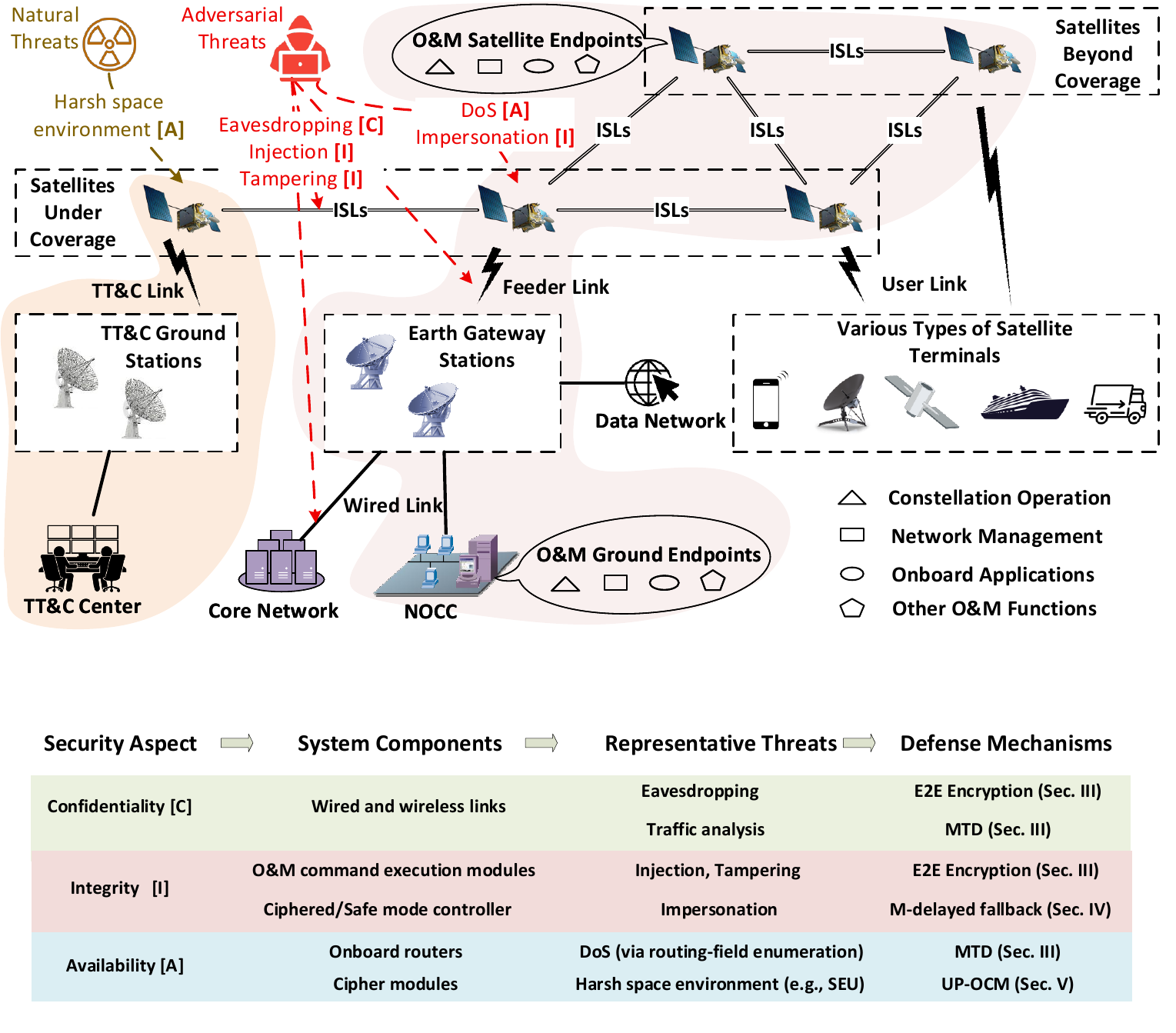}
\caption{ 
System architecture of \textcolor{black}{the considered} communication satellite constellation, where the TT\&C and O\&M functional layers are highlighted in yellow and pink, respectively. 
\textcolor{black}{The figure further provides a structured view of the threat model under both natural and adversarial conditions, and explicitly illustrates the mapping between security aspects (CIA), system components, representative threats, and the corresponding defense mechanisms. Ground and satellite O\&M endpoints are distinguished by different shapes.}
}
\label{fig:ScenarioV2}
\end{figure*}

Fig. \ref{fig:ScenarioV2} illustrates the architecture of the considered communication satellite constellation, \textcolor{black}{featuring inter-satellite links (ISLs) and onboard packet switching. 
Following the overall structure illustrated in Fig. \ref{fig:PaperStructure}, the functional layers of TT\&C and O\&M are supported by different links (namely TT\&C links and feeder links) and ground stations \cite{modenini2023tutorial}, and are typically operated by two separate entities.} 
The TT\&C Center is responsible for processing housekeeping telemetry and ensuring the safe operation of the satellite platform. 
In contrast, the Network Operations Control Center (NOCC) processes both housekeeping and payload telemetry, and is responsible for constellation operation, network management, onboard applications, and other emerging O\&M functions. 
The ground and satellite endpoints of these O\&M functions are illustrated with different shapes in Fig. \ref{fig:ScenarioV2}. 

\textcolor{black}{By leveraging ISLs and onboard packet switching, O\&M data (e.g., user-beam pointing management commands, routing configuration, and payload mode changes) can be relayed across satellites beyond the coverage of Earth gateway stations. 
This capability constitutes constellation-level O\&M and forms the foundation for continuous global communication services (i.e., the top network layer in Fig. \ref{fig:PaperStructure}) for moving user terminals, even when serving satellites lack direct feeder-link connectivity to ground gateway stations.}  

Following Section \ref{sec:Intro}, this article focuses on ensuring the CIA triad for constellation O\&M data flows between the ground endpoints in the NOCC and their corresponding satellite endpoints. 
\textcolor{black}{Fig.~\ref{fig:ScenarioV2} provides a structured view of the system architecture and explicitly illustrates the mapping between security aspects, system components, representative threats, and the proposed defense mechanisms.  
In particular, the figure highlights both natural and adversarial threats (shown in brown and red, respectively), with the corresponding CIA aspects marked by [C], [I], and [A].} 

\textcolor{black}{For confidentiality, the primary vulnerable components are the territorial wired links and wireless links\footnote{Intra-satellite data channels are assumed to be trusted under a secure supply chain without hardware backdoors.}, where adversaries may eavesdrop on sensitive data transmitted over these channels. Such risks are typically mitigated by E2E encryption. However, even when the data content is encrypted, advanced attackers may still infer behavioral information through traffic analysis, such as communication patterns and routing correlations. This residual leakage may still compromise confidentiality, thereby motivating the MTD mechanism proposed in Section~\ref{sec:1}.}

\textcolor{black}{For integrity, the primary targeted components include the onboard O\&M command execution modules, where adversaries may inject forged commands or modify legitimate messages in transit. These risks are typically mitigated by message authentication mechanisms integrated with E2E encryption. 
In addition, the ciphered/safe-mode controller, by design, must accept mode-transition commands to ensure system recoverability, which exposes it to impersonation attacks. This motivates the M-delayed fallback mechanism proposed in Section~\ref{sec:2}.} 

\textcolor{black}{For availability, the onboard routers responsible for real-time packet switching are vulnerable to DoS attacks via routing-field enumeration, as illustrated in Section~\ref{sec:1}. 
Such attacks can exhaust onboard processing resources and degrade system availability, motivating the use of the MTD mechanism.
In addition, onboard cipher modules are susceptible to harsh space environments, including Single Event Upsets (SEUs) and Single Event Latchup (SEL), which may impair or disable critical security functions. This motivates the UP-OCM mechanism proposed in Section~\ref{sec:3}.} 


\section{\textcolor{black}{Hybrid Protection Scheme} for Constellation O\&M: Cryptography Meets MTD}
\label{sec:1}

\textcolor{black}{Section \ref{sub:1-0} provides an overview of combining cryptography with MTD to form the hybrid protection scheme. Since the cryptography remains unchanged from the baseline design, the subsequent analysis focuses on MTD, with Sections \ref{sub:1-1}, \ref{sub:1-2}, and \ref{sub:1-3} addressing the procedures, effectiveness, and overhead, respectively.}


\subsection{\textcolor{black}{Overview of the Hybrid Protection Scheme}}
\label{sub:1-0}

\begin{figure}[t]
\centering
\includegraphics[width=1 \linewidth]{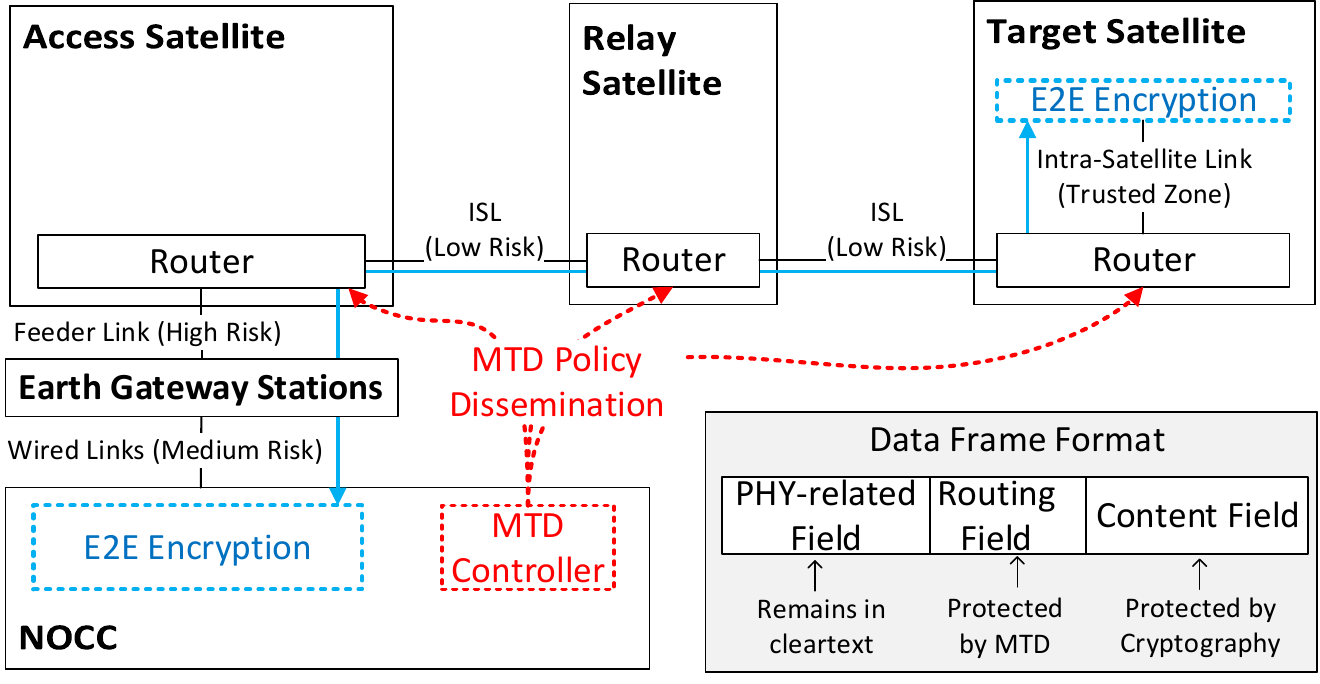}
\caption{ 
A typical O\&M data flow across satellites \textcolor{black}{under the proposed defense scheme.} The \textit{content field} of high \textcolor{black}{semantic sensitivity} is protected by E2E encryption between the NOCC and the target satellite; the \textit{routing field} of medium sensitivity is safeguarded using MTD; and the \textcolor{black}{\textit{PHY-related field}, which is operationally indispensable and carries no semantic sensitivity, remains in cleartext}. 
}
\label{fig:MTD}
\end{figure}

Fig. \ref{fig:MTD} illustrates a typical O\&M data flow across satellites, traversing territorial wired links, feeder links, and ISLs. 
Based on the \textcolor{black}{threat model in Section \ref{sec:System Architecture}, two encryption} schemes can be applied to protect O\&M data frames in a networked system such as a satellite constellation \cite{book2019application}. 
The first is H2H encryption, in which the encrypted data generated by the sending endpoint is decrypted and re-encrypted at every trusted relay node (i.e., the access satellite and relay satellites) until it reaches the receiving endpoint (i.e., the target satellite or the NOCC). 
The second is E2E encryption, in which the data frame is encrypted and decrypted only at the two endpoints.

For H2H encryption, even if \textcolor{black}{repeated decryption and re-encryption at all relay satellites} are omitted owing to the relatively low risk of ISLs\footnote{\textcolor{black}{Attacks on ISLs are assumed to be highly constrained due to stringent pointing, tracking, and interception requirements.}}, load imbalance still arises: the access satellite in Fig. \ref{fig:MTD} must encrypt and decrypt all \textcolor{black}{feeder-link traffic carrying constellation-level O\&M and user-related data frames}, thereby imposing a processing burden that scales with the constellation size. 
For a vanilla E2E encryption of the entire data frame (e.g., the NASA bulk security standard for spacecraft communication \cite{NASA_GSFC_STD_8012_2024}), the routers at relay and access satellites cannot decrypt the frame and thus cannot perform onboard routing under dynamic topologies, thereby violating the availability requirement. 

To this end, we propose a hybrid protection scheme that divides each frame into \textcolor{black}{Physical-layer-related (PHY-related) field}, routing field, and content field, as shown in Fig. \ref{fig:MTD}. 
\textcolor{black}{The PHY–related field, including the predefined synchronization sequences, the Modulation and Coding (MODCOD) parameters, and the Cyclic Redundancy Check (CRC) trailers, remain in cleartext. 
This design choice is consistent with common data frame constructions defined in existing standards, such as the Transmission Security (TRANSEC) implementations specified in Digital Video Broadcasting (DVB) standards \cite{dvb_a1554r1}.}  
The content field, due to its high sensitivity, employs E2E encryption between the target satellite and the NOCC, as illustrated by the blue arrow in Fig. \ref{fig:MTD}. 
The routing field\textcolor{black}{—introduced by advanced onboard routing functions under constellation-level O\&M—}leverages the concept of MTD to increase attacker uncertainty through dynamic system reconfigurations. 
Specifically, we employ a spatio-temporal MTD that combines \textit{per-hop replacement} with \textit{dynamic updating}, ensuring that the same routing-field value conveys different meanings across onboard routers and over time.  

\subsection{Operational Procedures of the Spatio-Temporal MTD}
\label{sub:1-1}
The red dashed arrows in Fig. \ref{fig:MTD} depict the operational flow of the proposed spatio-temporal MTD among its participants, namely the MTD controller at the NOCC and the onboard routers of the access, relay, and target satellites.

The MTD controller determines the policy and asynchronously disseminates it to selected \textcolor{black}{routers along the data transmission path}, thereby updating their interpretation of routing-field values. This dissemination is protected by E2E encryption between the NOCC and the satellites hosting each router, with the updated interpretation embedded in the encrypted content fields.

Each selected router then activates the new interpretation and applies it to real-time data processing. During the transition, old and new interpretations may temporarily coexist to prevent frame loss. Once all routers along the transmission path have switched to the new interpretation and no data frames depend on the old one, the outdated interpretation is deactivated, either automatically by timeout or under the MTD controller’s command. 
The following example illustrates the operational procedures of the spatio-temporal MTD. 

\begin{example} 
\label{example:MTD}
Consider a data frame traversing routers A, B, and C. Initially, it arrives at A with a routing field of 0x01.
According to its current interpretation, A replaces 0x01 with 0x03 and forwards the frame to B, which maps 0x03 to 0x05 before sending it to C.
This illustrates the \textit{per-hop replacement}.

When the MTD controller issues a \textit{dynamic update}, it disseminates a new interpretation to A and B, redefining the field value between them from 0x03 to 0x02.
Thereafter, A maps 0x01→0x02 and forwards the frame to B, which maps 0x02→0x05 before sending it to C.

During the switchover, router B temporarily retains both interpretations, since router A may not update simultaneously or an update may fail, causing it to still forward frames with 0x03. In such cases, B continues mapping 0x03→0x05 to prevent frame loss. 
\end{example}

\subsection{\textcolor{black}{Effectiveness against Different Attack Types}}
\label{sub:1-2}
To illustrate its effectiveness, we first identify the attack types and risks that emerge when the spatio-temporal MTD is not employed, while E2E encryption of the content field is still enforced.
\begin{itemize}
    \item \textbf{Transmission-Path Tracing (TPT) attacks:} By correlating routing-field observations of a data frame across different links (e.g., eavesdropping on both the downward feeder link and the upward user link), TPT attackers can reconstruct the full forwarding trajectory of a flow, thereby compromising routing anonymity, enabling selective disruption, and leaking operational patterns. 
    
    \item \textbf{Traffic Analysis (TA) attacks}: By conducting long-term observations of data frames (e.g., eavesdropping on the feeder downlink), TA attackers \cite{CNSSI4009} can accumulate sufficient samples for statistical analysis, thereby inferring sensitive communication patterns. 

    \item \textbf{Denial-of-Service (DoS) via routing-field enumeration:}  A DoS attacker can enumerate an $n$-bit routing field, which admits $2^n$ possible values, and inject frames covering this space. When aided by TA intelligence to identify likely-active links (spatial dimension) and favorable time windows (temporal dimension, e.g., peak traffic periods, scheduled transmissions, or ground-station contact windows), such an attack can flood routers and the target satellite with plausibly routed frames. Even if the payload is invalid, these frames still trigger onboard processing (e.g., parsing, routing lookup, authentication checks, telecommand handling), consuming critical resources and resulting in effective DoS despite strong content-field encryption. 
\end{itemize}

On the one hand, the \textit{per-hop replacement} mechanism thwarts TPT attacks by rewriting the routing-field value at each hop, so that an intercepted field at any single link does not reveal the \textcolor{black}{entire} forwarding trajectory, as illustrated in Example \ref{example:MTD}.  

On the other hand, \textit{dynamic updating} periodically reassigns the mapping between field values and forwarding semantics, making it difficult for an adversary who captures a field value over the air to interpret that value correctly and thereby undermining TA attacks. 
Moreover, DoS attacks based on brute-forcing the routing field are curtailed: attackers probing the value space after mappings have shifted will generate frames with stale or unmatched values, which the access router can detect and discard, thereby preventing resource exhaustion in downstream routers and target processing units.

\subsection{\textcolor{black}{Latency, Computational, and Communication Overhead}} 
\label{sub:1-3}

\textcolor{black}{The spatio-temporal MTD mechanism is transparent to real-time O\&M data processing (i.e., onboard routers process data frames using the same table-lookup procedures as in the baseline design), introducing no per-packet latency, processing complexity, or communication overhead during O\&M data delivery. 
The associated overhead is confined to the update phase, when the MTD controller generates and disseminates updated interpretation (field-mapping) tables to onboard routers.} 

\textcolor{black}{During the update phase, onboard processing is limited to simple flow-table replacement operations, as illustrated in Example \ref{example:MTD}. 
Compared to terrestrial network solutions that rely on per-hop MAC verification \cite{legner2020epic} to mitigate DoS attacks and path manipulation, our proposed approach significantly reduces onboard computational complexity. 
In addition, the data transmitted over the SGLs during updates consists only of compact flow-table information rather than per-packet authentication data. Moreover, since the design supports asynchronous updates, these update messages can be delivered as background traffic and scheduled opportunistically when sufficient bandwidth is available.}

\textcolor{black}{Importantly, all update-related computations on each satellite are strictly node-local and independent of the overall constellation size; therefore, the onboard computational burden does not increase as the constellation scales. The ground-based MTD controller handles update generation and coordination for multiple satellites, incurring a workload that grows linearly with the constellation size. This workload is absorbed by ground infrastructure with ample computational resources and thus does not impact onboard operations.}

\textcolor{black}{Despite the low overhead, achieving higher effectiveness against attacks typically requires more frequent updates, and a systematic exploration of the trade-off between overhead and effectiveness is left for future work.} 

\section{Ciphered-Mode and Safe-Mode Management: Toward Timely and Unexploitable Fallback} 
\label{sec:2}

\begin{figure*}[th]
\centering
\includegraphics[width=.9 \linewidth]{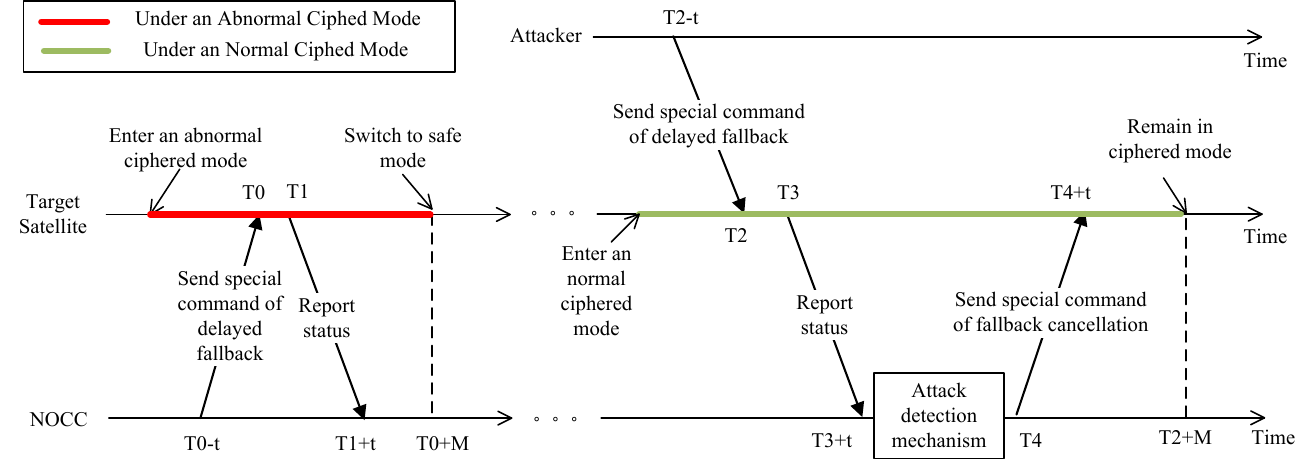}
\caption{ 
The proposed $M$-delayed fallback mechanism for both normal and abnormal ciphered modes. A fallback command opens an $M$-length window, allowing the NOCC’s attack detection mechanism to cancel forged requests; only if no cancellation telecommand arrives does the satellite switch to safe mode. 
}
\label{fig:delayedfallback}
\end{figure*}

To analyze recovery from \textcolor{black}{worst-case failures} of cipher functions, it is necessary to first define two typical satellite operational modes. 

In the \textit{ciphered mode} (or secure mode in the CCSDS standard \cite{book2019application}), the constellation O\&M is cryptographically protected, generating and accepting only encrypted and authenticated data. 
In contrast, the \textit{safe mode} emphasizes robustness of O\&M rather than security, and may follow different operational logics. For instance, the system may fall back to a default pre-set key \cite{NASA_GSFC_STD_8012_2024} instead of a dynamically updated one, or even degrade to a clear/transparent mode \cite{book2019application}, in which plaintext data are generated and accepted without authentication\footnote{Cryptographically unprotected does not mean defenseless. For example, embedding timestamps or sequence numbers into each data frame allows the onboard command receiver to validate freshness, accepting only timely or sequential frames. These lightweight checks make it far harder for attackers without protocol knowledge to replay intercepted uplink signals and disrupt constellation operations.}. 
To guard against malicious exploitation, transitions from ciphered mode to safe mode are enforced through cryptographic protection \cite{book2019application}. 

\subsection{Deadlock in Abnormal Ciphered Mode: Two Conventional Fallback Mechanisms} 
\label{sub:2-1}
Unlike ground deployment, onboard cipher modules must withstand harsh space conditions such as high-energy particle radiation and sub-optimal thermal control, which substantially increase the likelihood of faults.
When failures in the cipher modules lead to a complete malfunction of cipher functions—for example, due to the exhaustion of all backups or errors in module interface calls—uplinked constellation operation commands can no longer be decrypted or executed. 
Because in-orbit repair is impossible and remote recovery itself depends on the integrity of those teleoperation commands, the system falls into a circular dependency—a classic \textit{chicken-and-egg problem}—that can leave it locked in an abnormal ciphered mode. 


To break this deadlock, satellites in practice can rely on two fallback mechanisms. 
The first is a \textit{failsafe timeout}: the automatic fallback to the safe mode when no valid constellation command is received within a preset window. 
A classical example is the \textit{TC timer expiration} specified in the CCSDS standard \cite{book2019application}. 
The second is a \textit{backdoor command}: a plaintext and unauthenticated telecommand that triggers an immediate fallback to safe mode, which the satellite continues to accept even while operating in ciphered mode. 
The failsafe timeout preserves security, since regular encrypted uploads from NOCC prevent abuse. \textcolor{black}{However, the long delay before fallback leaves satellites exposed to risks ranging from attitude disruption to complete loss of contact.} 
The backdoor command enables immediate fallback; however, once disclosed, it can be exploited by attackers to force a transition into safe mode, and is therefore not recommended by the CCSDS standard \cite{book2019application}. 
We term this exploitation the \textit{Safe-Mode Downgrade (SMD) Attack}, an instance of the broader MITRE category `\textit{Exploit Reduced Protections During Safe Mode}' \cite{peled2023evaluating}. 

\subsection{Unified M-Delayed Fallback Mechanism} 
\label{sub:2-2}
In short, failsafe timeout guarantees security but weakens resilience, while backdoor command ensures resilience at the expense of security. 
To strike a balance, we propose a \textit{$M$-delayed fallback mechanism} with a delay window $M$, an attack detection mechanism in the NOCC, and two special commands of delayed fallback and fallback cancellation. 
Fig. \ref{fig:delayedfallback} illustrates this $M$-delayed fallback mechanism under abnormal and normal ciphered modes, indicated by the red and green lines, respectively.
In the abnormal case, NOCC sends a delayed fallback at $T0-t$, received at $T0$, leading the satellite to switch to plaintext at $T0+M$. In the normal case, a forged command injected at $T2$ triggers a status report at $T3$; since NOCC never issued such a command, the detection system flags a SMD attack and cancels the fallback, keeping the satellite in ciphertext at $T2+M$. 

The $M$-delayed fallback mechanism provides a unified representation of two practical fallback strategies, augmented with attack detection and a fallback-cancellation command at the NOCC. When $M=0$, it reduces to the \textit{backdoor command}; when $M$ equals the preset window, it corresponds to the \textit{failsafe timeout}. 
Based on the flowchart in Fig. \ref{fig:delayedfallback}, we observe the above trade-off role of $M$, which can be formally captured by the metrics of \textit{timeliness} and \textit{exploitability} defined below. 
\begin{definition}
\label{def:Timeliness}
Timeliness is the elapsed time from the onset of an abnormal ciphered mode until the system successfully transitions to plaintext.
\end{definition}

\begin{definition}
\label{def:Exploitability}
Exploitability is the likelihood that, while in a normal ciphered mode, an attacker can compromise constellation operations by exploiting the fallback mechanism.
\end{definition}

\textcolor{black}{A larger $M$ provides the ground segment with more opportunities to detect SMD attacks under normal ciphered operation, thereby reducing exploitability.} However, this comes at the cost of lower timeliness under abnormal ciphered modes. The proposed metrics thus provide a quantitative basis for assessing the resilience and security of constellation operation commands, and can be used to guide the optimal choice of $M$ in practice. 


The rationale behind the proposed $M$-delayed fallback mechanism is that, in satellite scenarios, forging messages is relatively easy, whereas blocking regular ones is much harder. 
\textcolor{black}{Consequently, coordinated attacks that inject forged delayed-fallback commands while simultaneously blocking both the status-report telemetry and the NOCC’s cancellation command in Fig.~\ref{fig:delayedfallback} throughout the entire M-delay window are highly unlikely to succeed. A more realistic threat is that} an advanced SMD attacker may persistently inject forged delayed-fallback commands, forcing the NOCC to periodically send cancellation commands to prevent unintended switches from normal ciphered mode to safe mode. In such cases, the NOCC can further modify the command source code or format of the delayed-fallback instruction, compelling the attacker to expend significantly greater effort to discover the new command. While the detailed design is left for future work, it is clear that this setting produces a defensive game biased in favor of the defender, in contrast to most conventional security scenarios where attackers typically hold the advantage. 

\section{From Endpoint Invocation to Unified Pooling: Enhancing Cipher Module Reliability} 
\label{sec:3}

Although Section \ref{sec:2} describes system-level fallback mechanisms for \textcolor{black}{worst-case failures} of onboard cipher functions, module-level redundancy within the encryption subsystem is generally expected to make such activations rare. 
To make the module-level backup compatible with the fallback mechanism, invocation-based encryption is used so that O\&M satellite endpoints can bypass their cipher modules and switch from ciphered mode to safe mode. 

\begin{figure}[t]
\centering
\includegraphics[width=1 \linewidth]{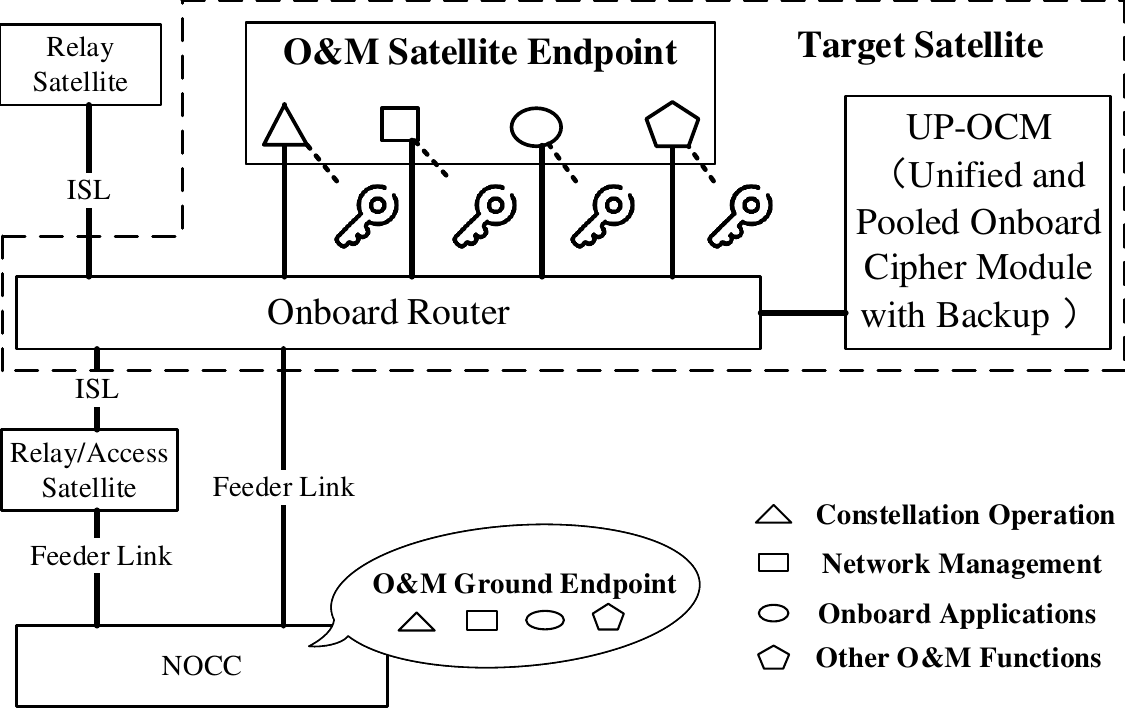}
\caption{ 
Revised invocation-based encryption scheme: onboard routers invoke the security function for encryption and decryption via the UP-OCM, which maintains a shared resource pool of cipher module backups for all O\&M functions. 
}
\label{fig:backup}
\end{figure}

In a typical invocation-based encryption setup, O\&M satellite endpoints interact directly with their cipher modules, represented by the key icon in Fig. \ref{fig:backup}. 
Specifically, each endpoint submits plaintext (or ciphertext) data to the cipher module for encryption (or decryption) and receives the corresponding ciphertext (or plaintext) output. 
However, this setup scales poorly: the number of cipher modules grows with the number of endpoints, each requiring its own dedicated backup, with sharing across endpoints generally infeasible. 

Unification and resource pooling \cite{fuchs2018dynamic} could mitigate this inefficiency. Yet, because endpoints are usually deployed on physically independent machines, consolidating cipher modules into a single unit would necessitate cross-machine invocations, increasing interconnect complexity and reducing link reliability. 

\subsection{\textcolor{black}{Security-Aware} Routers for Cipher Module Pooling} 
To this end, we propose a revised invocation-based encryption scheme tailored to constellations with onboard routers.
As shown in Fig. \ref{fig:backup}, we introduce a new independent module, the Unified and Pooled Onboard Cipher Module with Backup (UP-OCM), which interacts directly with the onboard router rather than with individual endpoints.

In this setting, the onboard router embeds the following security functions to manage the ciphered and safe modes introduced in Section \ref{sec:2}: 
\begin{itemize}
    \item When the target satellite operates in safe mode, the router bypasses the UP-OCM and carries out its native role of routing. 
    \item When the target satellite enters ciphered mode, its router distinguishes traffic flows using the unencrypted routing field defined in Section \ref{sub:1-0}. 
    Data frames of relay and access satellites are routed directly, whereas O\&M data frames of the target satellite are processed in coordination with the UP-OCM for encryption or decryption. 
    \item The router acts as the responsible entity for the $M$-delayed fallback mechanism described in Section \ref{sub:2-2}, ensuring a controlled transition from ciphered to safe mode when the UP-OCM is abnormal. 
\end{itemize}

The rationale for enabling the security function in onboard routers is twofold. First, as illustrated in Fig. \ref{fig:MTD}, an O\&M data frame enters its trusted zone upon arrival at the target satellite. Hence, decryption can be performed at the onboard router (via the UP-OCM), after which the frame is delivered in plaintext to the O\&M endpoints. Second, the onboard router acts as a centralized processing node through which all O\&M data frames of the target satellite must pass for routing, thereby eliminating the possibility of bypass. 

\textcolor{black}{Nevertheless, the proposed architecture is developed under a clearly defined threat model in Section \ref{sec:System Architecture}. If the target satellite itself is fully compromised, the trusted-zone assumption no longer holds, and intra-satellite communications may be exposed. Moreover, a transition toward distributed onboard routing would fundamentally alter the attack surface and protection granularity. Such scenarios correspond to a different threat regime and require complementary architectural defenses, which can be investigated in future work.}

\subsection{Simulation-Based Analysis of Reliability Enhancement} 
Since the UP-OCM can encrypt and decrypt all O\&M flows in a unified manner while applying distinct cryptographic keys to different flows, it is able to maintain a shared pool of backup units accessible across flows. 
With the same SWaP overhead, this design revision enhances overall reliability, as demonstrated by the following simulation results. 
\textcolor{black}{In the simulation, SWaP overhead is abstracted by the number of backup cipher modules, without relying on implementation-specific hardware parameters, thereby capturing the fundamental reliability and scalability trends of the proposed design.} 

In the typical invocation-based encryption scheme with $N_1=3$ endpoints \textcolor{black}{(corresponding to $N_1$ O\&M flows)}, each endpoint is provisioned with $N_2$ cipher modules: one active and $N_2-1$ backups.
In contrast, our revised scheme leverages pooling, whereby the UP-OCM centralizes $N_1 \times N_2 = 6$ cipher units in a shared pool that can be flexibly allocated across \textcolor{black}{$N_1$ O\&M flows} as active or backup. 
It is straightforward that the revised scheme reduces to the typical scheme when $N_2 = 1$ or $N_1 = 1$. 

Following \cite{grile2025statistical}, we model the lifetime of each cipher module/unit (either active or backup) using an i.i.d. Weibull distribution with shape parameter $\beta=0.6$ and scale parameter $\eta=10$, with time expressed in years. 
\textcolor{black}{As a standard reliability model for electronic systems, the Weibull distribution is adopted here to capture relative reliability trends rather than exact failure times.} 
The constellation O\&M is defined to be in normal operation only when all its constituent O\&M functions are operational; each function, in turn, remains operational as long as at least one of its cipher modules is functional. 
Therefore, under the revised setting, the constellation O\&M operates normally only if at least $N_1$ out of the total $N_1 \times N_2$ cipher modules remain functional. 

\begin{figure}[t]
\centering 
\begin{subfigure}{0.45\textwidth}
\includegraphics[width=1\columnwidth,height=.8\columnwidth]{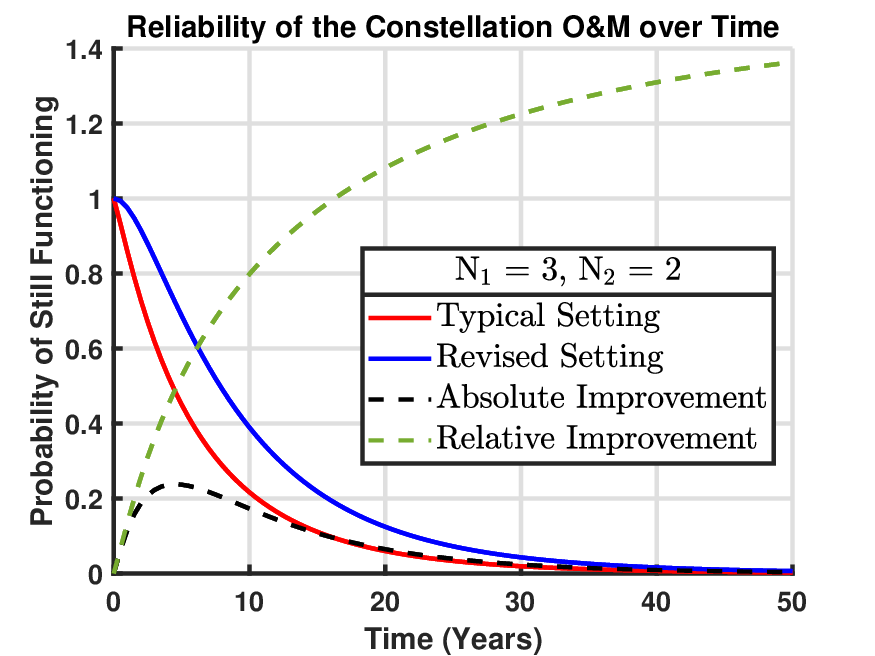} 
\caption{ 
Enhanced reliability under the revised setting. 
}
\label{fig:distribution}
\end{subfigure}\hfil 
\begin{subfigure}{0.45\textwidth}
\includegraphics[width=1\columnwidth,height=.8\columnwidth]{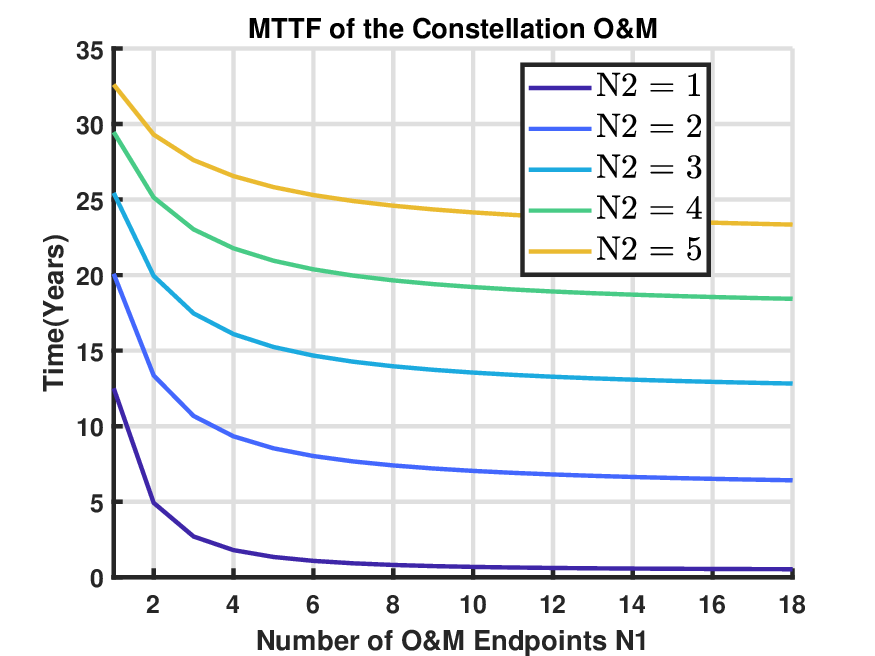} 
\caption{ 
MTTF of the Constellation O\&M Function. 
}
\label{fig:MTTF}
\end{subfigure}
\caption{Simulation results under the Weibull distribution with shape parameter $\beta=0.6$ and scale parameter $\eta=10$. 
\label{fig:simulation results}
}
\end{figure}

Fig. \ref{fig:simulation results} illustrates the corresponding simulation results.
As shown in Fig.~\ref{fig:distribution} for $N_1 = 3$ and $N_2 = 2$, the revised setting (blue line) markedly improves the reliability of the constellation O\&M. 
Reliability, shown on the y-axis, is defined as the probability that all O\&M functions remain operational until the time indicated on the x-axis. 
Compared with the typical setting (red line), the absolute improvement (black dashed line) peaks at $0.23$ around year $4.5$ before declining, while the relative improvement (green dashed line) increases monotonically to approximately $140\%$ over $50$ years of operation. 
Fig.~\ref{fig:MTTF} shows the MTTF of the constellation O\&M under the revised setting. 
On the one hand, for a given $N_1$, increasing the number of backups steadily enhances reliability, as expected from the added redundancy. 
The improvement is more evident for smaller system scales (i.e., when $N_1$ is small) or lower backup degrees (i.e., when $N_2$ is small). 
On the other hand, for a fixed $N_2$, the overall reliability decreases as the system scales up, since the failure of any O\&M function leads to the failure of the entire constellation. 
\textcolor{black}{However, as the number of endpoints continues to increase, the rate of the MTTF degradation becomes progressively slower and eventually converges to a steady value. This behavior demonstrates the scalability of UP-OCM under increasing O\&M flows.}


\section{Conclusion}
\label{sec:conclusion}


\textcolor{black}{This work investigates the challenges of securing constellation-level O\&M under both normal operation and fault-response scenarios, and proposes a holistic protection framework characterized by the following key properties.}

\textcolor{black}{The proposed mechanisms in the framework are orthogonal and composable, enabling independent or joint deployment without introducing new failure modes. 
The framework mitigates both internal system faults and external adversarial threats, including coordinated attacks that may occur sequentially or simultaneously. 
It is compatible with existing CCSDS and DVB standards and represents a practical refinement that can be readily integrated into current constellation architectures. 
In addition, the design emphasizes generality and configurability, allowing operators to select key parameters (e.g., the fallback window size) according to constellation-specific characteristics such as orbital period and required security level. Finally, the framework demonstrates good scalability to large-scale constellations.}

\textcolor{black}{In conclusion, this work offers a practical roadmap for securing constellation-level O\&M through a holistic integration of efficiency, resilience, and reliability.}

\bibliographystyle{IEEEtran}
\bibliography{RSE}

\end{document}